\documentclass[twocolumn,showpacs,preprintnumbers,amsmath,amssymb]{revtex4}
\usepackage{amsmath,amsfonts,amssymb,graphics,graphicx,epsfig,color,times,bbm}
\usepackage{esint}
\usepackage{dcolumn}% Align table columns on decimal point
\begin{document}

\preprint{APS/123-QED}

\title{Dipole-dipole shift of quantum emitters\\ coupled to surface plasmons of a nano-wire
}% Force line breaks with \\

\author{David Dzsotjan$^{1, 2}$}
 %\altaffiliation[Also at ]{}%Lines break automatically or can be forced with \\
\author{J\"{u}rgen K\"{a}stel$^{1}$\footnote{Now at German Aerospace Center (DLR), Institute of Technical Physics, Stuttgart, Germany}}
\author{Michael Fleischhauer$^1$}
\affiliation{$^{1}$ Department of Physics and research center OPTIMAS, University of Kaiserslautern, Germany;\\ $^{2}$ Research Institute for Particle and Nuclear Physics, H-1525 Budapest, Hungary}


\date{\today}% It is always \today, today,

\begin{abstract}
Placing quantum emitters close to a metallic nano-wire, an effective interaction can be achieved over distances large compared to the resonance wavelength due to the strong coupling between emitters and the surface plasmon modes of the wire. This leads to modified collective decay rates, as well as to Lamb and dipole-dipole shifts. In this paper we present a general method for calculating these level shifts which we subsequently apply to the case of a pair of atoms coupled to the guided modes of a nano-wire.
\end{abstract}

\pacs{Valid PACS appear here}% PACS, the Physics and Astronomy
\maketitle

%%%%%%%%%%%%%%%%%%%%%%%%%%%%%%%%%%%%%%%%%%%%%%%%%%%%%%%%%%%%%%%%%%%%%%%%%%%%%%%

\section{Introduction}

Coupling quantum systems to a common reservoir results in an effective
interaction between these systems. In the case of spontaneous emission of light from quantum emitters this manifests itself in phenomena such as superradiance and dipole-dipole interactions. In free space these
effects quickly disappear as soon as the average distance between the emitters 
exceeds the resonant wavelength. The situation changes dramatically, 
however, if the most dominant reservoir modes are reduced to one or zero spatial dimensions as in the case of a nano-wire or a single-mode resonator. Here interactions over large distances can emerge. Both the collective decay rates as well as the effective 
interaction Hamiltonian are fully determined by the dyadic Green's function of the electromagnetic field characterizing the response of the tailored reservoir.
Calculating the collective decay rates requires to determine the Green's tensor at a given frequency, typically the resonance frequency of the involved emitters.
However, in order to determine the level shifts one must perform a principal value integral over the whole positive frequency spectrum which raises serious additional difficulties.

In this paper we present a method to simplify the calculation of the Lamb and dipole-dipole shifts. One of the key steps is to introduce imaginary frequencies and extend the integrals over frequency into the complex plane. There are also other situations when introducing imaginary frequencies and using complex frequency integrals helps to simplify the original problem, for example in case of calculating Casimir-Polder potentials and investigating van der Waals interactions (\cite{Buhmann-JOptB-2004}, \cite{Buhmann-PRA-2004}, \cite{Buhmann-ProgQED-2007} and references therein). 
We also introduce a special Kramers-Kronig relation which, combined with the aforesaid integral extension lets us transform the original expression for the level shifts into a much more convenient form.

We apply the method to a particular example where a pair of quantum emitters are coupled to the surface plasmon modes of a nano-wire. This system is interesting because it enables to attain strong atom-field coupling and single-site addressability at the same time. Because of the small transverse mode area of the plasmons of a metallic cylinder with a sub-wavelength radius, a strong Purcell effect arises between the plasmons and a single emitter placed close to the wire (\cite{Chang-PRL-2006}, \cite{Chang-PRB-2007}, \cite{Dzsotjan-PRB-2010}). The effect of strong coupling to the plasmon modes of a waveguide has been studied for various specific scenarios (\cite{Chen-OptExp-2010}, \cite{Martin-Cano-NanoLett-2010}). The system of a single emitter coupled to a wire has been proposed as an efficient single-photon generator (\cite{Chang-PRL-2006}), as well as a single-photon transistor (\cite{Chang-NatPhys-2007}) and the coupling has been experimentally demonstrated (\cite{Akimov-Nature-2007}).
 Having a pair of emitters coupled to the guided modes, we derived an inter-emitter distance dependent superradiance effect in \cite{Dzsotjan-PRB-2010}. Here we calculate the dipole-dipole level shifts of the two-atom system.

\section{General method}\label{sec:general_method}

Let us consider a system of $N$ two-level quantum emitters, characterized by the
lowering and raising operators $\hat \sigma_j$ and $\hat \sigma_j^\dagger$ coupled to
a common electromagnetic reservoir. Under conditions that permit the dipole-, rotating wave-,  and Markov approximation tracing out the reservoir modes leads to
a master equation for the atoms
\begin{eqnarray}
 \dot{\hat{\rho}}=&\mathrm{i}\!\!\!\!&\sum_{m,n=1}^N\delta\omega_{mn}\left[\hat{\sigma}^\dagger_m\hat{\sigma}_n,\hat{\rho}\right]\nonumber\\
&-&\sum_{m,n=1}^N\frac{\Gamma_{mn}}{2}\left(\hat{\sigma}^\dagger_m\hat{\sigma}_n\hat{\rho}+\hat{\rho}\hat{\sigma}^\dagger_m\hat{\sigma}_n-2\hat{\sigma}_n\hat{\rho}\hat{\sigma}^\dagger_m\right).
\end{eqnarray}
The first, Hermitian term of the rhs contains the $\delta\omega_{nn}$ Lamb shifts and $\delta\omega_{m\ne n}$ dipole-dipole shifts. In the dissipative term we find the single-atom decay rates $\Gamma_{nn}$ and the inter-atomic decay couplings $\Gamma_{m\ne n}$. Their explicit form is:
\begin{eqnarray}
\Gamma_{mn}(\omega_A)&=&\frac{2\omega_A^2 d_{m_i} d_{n_j}}{\hbar\epsilon_0c^2}\mathrm{Im}\left[G_{ij}(\vec{r}_m,\vec{r}_n,\omega_A)\right] \label{eq:Gamma}\\
\delta\omega_{mn}(\omega_A)&=&\frac{d_{m_i}d_{n_j}}{\hbar\epsilon_0\pi}\mathbb{P}\!\!\int_{0}^{\infty}\!\!\!\!\!d\omega\frac{\omega^2}{c^2}\frac{\mathrm{Im}\left[G_{ij}(\vec{r}_m,\vec{r}_n,\omega)\right]}{\omega-\omega_A}.\nonumber\\
\label{eq:dipoleshift}
\end{eqnarray}
Here, $G_{ij}(\vec r_m,\vec r_n,\omega)$ is the $\{i,j\}$ component of the Green's tensor for the electromagnetic field including the interaction with
a passive medium such as a nano-wire, evaluated at frequency $\omega$ and at positions $\vec r_m$ and $\vec r_n$. It fulfills the Maxwell-Helmholtz wave equation
\begin{equation}
\left[\nabla\times\frac{1}{\mu(\vec{r},\omega)}\nabla\times-\frac{\omega^2}{c^2}\epsilon(\vec{r},\omega)\right]\bar{\bar{G}}(\vec{r},\vec{r}^\prime,\omega)=\bar{\bar{I}}\delta(\vec{r}-\vec{r}^\prime),
\end{equation}
with the proper boundary conditions. $\epsilon(\vec{r},\omega)$ and $\mu(\vec{r},\omega)$ are the relative electric permittivity and magnetic permeability, respectively. 
In case of transition frequencies $\omega_A$ for which the rotating wave approximation is valid and which are far from the ultraviolet domain, we can use the full (vacuum plus material part) Green's tensor for calculating $\Gamma_{mn}$. 
For calculating $\delta\omega_{mn}$ one has to perform however an integral over the whole spectrum. Since the atom-field coupling is treated in a non-relativistic 
way, this approach does not take into account the relativistic high-frequency contributions correctly. A well known consequence of this is that the vacuum level shifts (Lamb shift) obtained from (\ref{eq:dipoleshift}) are incorrect.
Here a rather involved calculation based on relativistic quantum field theory is
required, taking into account all possible transitions 
of the emitter and including proper renormalizations.
However, if we are interested only in the changes produced by the presence of a
medium, this problem can be circumvented. The medium tailors the reservoir modes
only within a certain frequency range and becomes transparent in the high-frequency domain. Thus, calculating the effects relative to the case of free-space vacuum, i.e using $\bar{\bar{G}}^{med}=\bar{\bar{G}}-\bar{\bar{G}}^{vac}$ instead of the full $\bar{\bar{G}}$, the above equations give correct expressions
also for the Lamb or dipole-dipole shifts relative to vacuum. In other words calculating only the material-induced level shift introduces an automatic renormalization and lets us get rid of the ultraviolet divergences. 

Deriving (\ref{eq:Gamma}) and (\ref{eq:dipoleshift}) we also used a Markov approximation. This is possible as long as the calculated decay rates $\Gamma_{mn}(\omega)$ and shifts $\delta\omega_{mn}(\omega)$ depend only slowly
on frequency $\omega$, i.e. do not change appreciably over frequency ranges of the
order of $\Gamma_{mn}$ and $\delta\omega_{mn}$. It should be kept in mind that even if the spectral response of the medium is flat, retardation effects can cause a
spectral dispersion of the Green's tensor at two different positions $\vec r_m$ and
$\vec r_n$ with a characteristic width given by $c/|\vec r_m-\vec r_n|$
\cite{Kaestel-PRA-2005}.

Even though ultraviolet divergencies are eliminated in (\ref{eq:dipoleshift}) by considering only the changes due to the medium, the expression still contains an integral which is rather difficult to calculate.
In the following, we will give a general method for simplifying this expression. For this we shall use a generalized Kramers-Kronig relation.

\subsection{Kramers-Kronig relations for the Green's tensor}\label{subsec:Kramers-Kronig_relations_for_the_Green's_tensor}

The full Green's tensor $\bar{\bar{G}}(\omega)$ does not have any poles on the upper complex half-plane because of causality. Thus Kramers-Kronig relations (see Appendix \ref{appendix:Kramers-Kronig_relation_generally}) apply, e.g.
\begin{eqnarray}\label{eq:Kramers-Kronig}
\mathrm{Re}\{\bar{\bar{G}}(\omega_A)\}&=&\frac{1}{\pi}\mathbb{P}\int_{-\infty}^{\infty}\!\!\!\! d\omega\frac{\mathrm{Im}\{\bar{\bar{G}}(\omega)\}}{\omega-\omega_A}.
\end{eqnarray}
Because the Green's function inherits the symmetry $\bar{\bar{G}}(-\omega^*)=\bar{\bar{G}}^*(\omega)$ from $\epsilon(\omega)$, equation (\ref{eq:Kramers-Kronig}) can also be written in the form
\begin{eqnarray}
 \mathrm{Re}\{\bar{\bar{G}}(\omega_A)\}&=&\frac{2}{\pi}\mathbb{P}\int_{0}^{\infty}\!\!\!\!
d\omega\frac{\omega\mathrm{Im}\{\bar{\bar{G}}(\omega)\}}{\omega^2-\omega_A^2}.
\end{eqnarray}
An important step in the derivation of the Kramers-Kronig relation is the integration over the semicircle ($\curvearrowleft$ contribution) in the complex upper half plane. 
As this integration is done for large $|\omega|$, using the vacuum Green's tensor (\cite{deVries-RMP-1998}) 
\begin{eqnarray}
 \bar{\bar{G}}^{vac}(\vec{r},\omega)=\frac{e^{\mathrm{i}\frac{\omega}{c}r}}{4\pi r}\Biggl[\left(1-\frac{1}{\mathrm{i}\frac{\omega}{c}r}-\frac{1}{\frac{\omega^2}{c^2}r^2}\right)\mathbbm{1}\nonumber\\
+\left(-1+\frac{3}{\mathrm{i}\frac{\omega}{c}r}+\frac{3}{\frac{\omega^2}{c^2}r^2}\right)\hat{r}\otimes\hat{r}\Biggr]+\frac{\delta(\vec{r})}{3\frac{\omega^2}{c^2}}\mathbbm{1}.
\end{eqnarray}
is a good approximation. At large $|\omega|$, it goes as
\begin{equation}
 \lim_{|\omega|\rightarrow\infty}\bar{\bar{G}}^{vac}(\vec{r},\omega)=\frac{e^{\mathrm{i}\frac{\omega}{c}r}}{4\pi r}\left(\mathbbm{1}-\hat{r}\otimes\hat{r}\right).
\end{equation}
To perform the $\curvearrowleft$ integral we use $\omega=|\omega|e^{i\alpha}$ and approach the two points where the $\curvearrowleft$ part joins the real axis ($\delta\leq\alpha\leq\pi-\delta$). We then find
\begin{eqnarray}
 I&=&|\omega|\int_\curvearrowleft \!\!\!
d(e^{\mathrm{i}\alpha})e^{\mathrm{i}\frac{r}{c}|\omega|e^{\mathrm{i}\alpha}}\nonumber\\
&=&|\omega|\int_{1+\mathrm{i}\delta}^{-1+\mathrm{i}\delta}\!\!\! d(e^{\mathrm{i}\alpha})e^{\mathrm{i}\frac{r}{c}|\omega|e^{\mathrm{i}\alpha}}\nonumber\\
&=&-\frac{2c}{r}\sin(\frac{|\omega|}{c}r)e^{-\frac{|\omega|}{c}r\delta}.
\end{eqnarray}
Taking $|\omega|$ to infinity and $\delta$ to zero, we get
\begin{equation}
 \lim_{\delta\rightarrow 0}\lim_{|\omega|\rightarrow\infty}I=0.
\end{equation}
Because the integrand on any point of the $\curvearrowleft$ contour part goes exponentially fast to $0$, the integral will vanish on this part of the contour even if we multiply the integrand with a polynomial of $\omega$. 
In particular we find that the variant of the Kramers-Kronig relation
\begin{eqnarray}\label{eq:Kramers-Kronig-special}
\frac{\omega_A^2}{c^2}\mathrm{Re}\{\bar{\bar{G}}(\omega_A)\}&=&\frac{2}{\pi}\mathbb{P}\int_{0}^{\infty}\!\!\!\!
d\omega\frac{\omega^2}{c^2}\frac{\omega\mathrm{Im}\{\bar{\bar{G}}(\omega)\}}{\omega^2-\omega_A^2}
\end{eqnarray}
holds as well. For the reasons stated above, (\ref{eq:Kramers-Kronig}) and (\ref{eq:Kramers-Kronig-special})  - being true for $\bar{\bar{G}}$ and $\bar{\bar{G}}^{vac}$ - are valid  for the material contribution  $\bar{\bar{G}}^{med}$ also.

\subsection{Medium contribution to the Lamb and dipole-dipole shift}\label{subsec:calculating_the_Lamb_and_dipole-dipole_shift}

We can rewrite the principal value integral in (\ref{eq:dipoleshift})  as
\begin{equation}\label{eq:princ_int}
 I_1=\mathbb{P}\int_0^\infty \!\!\!\!
d\omega\frac{\omega^2}{c^2}\frac{\mathrm{Im}\{G_{ij}^{med}(\vec{r}_m,
\vec{r}_n,\omega)\}}{\omega^2-\omega_A^2}(\omega+\omega_A),
\end{equation}
where we replaced the full $\bar{\bar{G}}$ by $\bar{\bar{G}}^{med}=\bar{\bar{G}}-\bar{\bar{G}}^{vac}$, i.e. the contribution of the medium. We can now substitute the variant of the Kramers-Kronig relation (\ref{eq:Kramers-Kronig-special}) into (\ref{eq:princ_int}) which yields:
\begin{eqnarray}\label{eq:princ_int2}
\begin{aligned}
 I_1=&\frac{\pi}{2}\frac{\omega_A^2}{c^2}\mathrm{Re}\{G_{ij}^{med}(\omega_A)\}\\
+&\mathrm{Im}\left\{\mathbb{P}\int_0^\infty\!\!\!\! d\omega\frac{\omega^2}{c^2}G_{ij}^{med}(\omega)\frac{\omega_A}{\omega^2-\omega_A^2}\right\}.
\end{aligned}
\end{eqnarray}
Now, we will try to eliminate the principal value from the second term of (\ref{eq:princ_int2}). To do this, we will have to transfer the integral from the real axis to the imaginary axis. This, however, shall have its advantages because the Green's tensor behaves much more smoothly for purely imaginary frequencies: oscillations become exponentially decreasing functions.

Resolving the principal value
\begin{eqnarray}\label{eq:princ_val}
\begin{aligned}
 \mathbb{P}\left(\frac{\omega_A}{\omega^2-\omega_A^2}\right)=\frac{1}{2}\biggl(&\frac{1}{\omega-\omega_A-i\delta}-\mathrm{i}\pi\delta(\omega-\omega_A)\\
-&\frac{1}{\omega+\omega_A-\mathrm{i}\delta}+\mathrm{i}\pi\delta(\omega+\omega_A)\biggr),
\end{aligned}
\end{eqnarray}
the second term in (\ref{eq:princ_int2}) assumes the form
\begin{eqnarray}\label{eq:princ_int3}
 I_2&&=-\frac{\pi}{2}\mathrm{Im}\left\{\mathrm{i}\frac{\omega_A^2}{c^2} 
G_{ij}^{med}(\omega_A)\right\}+\frac{1}{2}\mathrm{Im}\biggl\{\int_0^\infty \!\!\!\!d\omega\frac{\omega^2}{c^2}G_{ij}^{med}(\omega)\nonumber\\
&&\times\left(\frac{1}{\omega-\omega_A-\mathrm{i}\delta}-\frac{1}{\omega+\omega_A-\mathrm{i}\delta}\right)\biggr\}.
\end{eqnarray}
Because the integral goes from $0$ to $\infty$ on the real axis, we can create a closed contour in the upper right quarter of the complex plane.
The integral over the curved part ($\cal C$) of the contour will again disappear, so we can write
\begin{equation}
\int_0^\infty\,\rightarrow\,\varointctrclockwise-\int_{\cal C}-\int_{\mathrm{i}\infty}^0
=\varointctrclockwise-\int_{\mathrm{i}\infty}^0.
\end{equation}
Using $\omega=i\kappa$ on the imaginary axis, we get
\begin{eqnarray}\label{eq:princ_int4}
\begin{aligned}
 I_2&=\frac{\pi}{2}\frac{\omega_A^2}{c^2}\mathrm{Re}\{G_{ij}^{med}(\omega_A)\}-\mathrm{Im}\biggl\{\int_0^\infty\!\!\!\! d\kappa\frac{\kappa^2}{c^2}\frac{\mathrm{i}G_{ij}^{med}(\mathrm{i}\kappa)}{2}\\
&\times\left(\frac{1}{\mathrm{i}\kappa-\omega_A}-\frac{1}{\mathrm{i}\kappa+\omega_A}\right)\biggr\}.
\end{aligned}
\end{eqnarray}
Substituting (\ref{eq:princ_int4}) into (\ref{eq:princ_int2}) and then that into (\ref{eq:dipoleshift}), we get for the Lamb and dipole-dipole shift
\begin{eqnarray}\label{eq:dipoleshift_final}
\begin{aligned}
 \delta\omega_{mn}(\omega_A)=&\frac{d_{m_i}d_{n_j}}{\hbar\pi\epsilon_0}\biggl[\pi\frac{\omega_A^2}{c^2}\mathrm{Re}\{G_{ij}^{med}(\vec{r}_m,\vec{r}_n,\omega_A)\}\\
+&\int_0^\infty\!\!\!\! d\kappa\frac{\kappa^2}{c^2}\mathrm{Re}\{G_{ij}^{med}(\vec{r}_m,\vec{r}_n,\mathrm{i}\kappa)\}\frac{\omega_A}{\kappa^2+\omega_A^2}\biggr].
\end{aligned}
\end{eqnarray}
In this form we no longer have to worry about the principal value. As an additional benefit, transferring the integration to the imaginary axis makes the Green's tensor much better behaved (exponential decay instead of oscillations) which is very useful when calculating the shift by numerical means. 

\subsection{Origin of the integral term}\label{sec:origin_of_int_term}

The expression for the dipole-dipole shift found in the literature (for example, \cite{Chang-PRA-2004}, \cite{Gonzalez-Tudela-PRL-2011}, \cite{VanVlack-arxiv-2011}) usually involves only the first term of the right hand side of (\ref{eq:dipoleshift_final}) or, equivalently, the shift is proportional to the real part of the electric field at the atomic frequency. This is, however, an approximation that relies on the assumption that only frequencies around $\omega_A$ contribute to the dipole-dipole shift considerably. We can obtain this result if
we only keep the first two terms in (\ref{eq:princ_val}), arguing that the other terms have their greatest contribution at around $\omega=-\omega_A$ which is not contained by the region of integration in (\ref{eq:princ_int}). 
However, there is a fundamental reason why in general the integral term in (\ref{eq:dipoleshift_final}) must be present. Starting from (\ref{eq:princ_int2}), we can write
\begin{eqnarray}\label{eq:dipoleshift_integralterm1}
\begin{aligned}
 \mathbb{P}\int_0^\infty\!\!\!\!
 d\omega\frac{\omega^2}{c^2}\frac{\mathrm{Im}\{G_{ij}^{med}(\omega)\}}{\omega-\omega_
A}=\frac{\pi}{2}\frac{\omega_A^2}{c^2}\mathrm{Re}\{G_{ij}^{med}(\omega_A)\}&\\
+\omega_A\mathbb{P}\int_0^\infty\!\!\!\! d\omega\frac{\omega^2}{c^2}\frac{\mathrm{Im}\{G_{ij}^{med}(\omega)\}}{\omega^2-\omega_A^2}&.
\end{aligned}
\end{eqnarray}
Resolving the integral term on the rhs of Eq. (\ref{eq:dipoleshift_integralterm1}), we get two terms, one of which is identical to the one on the lhs of the equation. Thus, one obtains
\begin{eqnarray}\label{eq:dipoleshift_integralterm2}
 \begin{aligned}
  \mathbb{P}\int_0^\infty\!\!\!\! 
d\omega\frac{\omega^2}{c^2}\frac{\mathrm{Im}\{G_{ij}^{med}(\omega)\}}{\omega-\omega_A}=\pi\frac{\omega_A^2}{c^2}\mathrm{Re}\{G_{ij}^{med}(\omega_A)\}&\\
-\mathbb{P}\int_0^\infty\!\!\!\! d\omega\frac{\omega^2}{c^2}\frac{\mathrm{Im}\{G_{ij}^{med}(\omega)\}}{\omega+\omega_A}&.
 \end{aligned}
\end{eqnarray}
Changing the integration variable on the rhs of Eq. (\ref{eq:dipoleshift_integralterm2}) to $\tilde{\omega}=-\omega$ and using the fact that $\mathrm{Im}\{G_{ij}(\omega)\}$ is an odd function of $\omega$, we get
\begin{eqnarray}
 \begin{aligned}
  \mathbb{P}\int_0^\infty\!\!\!\! 
d\omega\frac{\omega^2}{c^2}\frac{\mathrm{Im}\{G_{ij}^{med}(\omega)\}}{\omega-\omega_A}=\pi\frac{\omega_A^2}{c^2}\mathrm{Re}\{G_{ij}^{med}(\omega_A)\}&\\
-\mathbb{P}\int_{-\infty}^0\!\!\!\! d\tilde{\omega}\frac{\tilde{\omega}^2}{c^2}\frac{\mathrm{Im}\{G_{ij}^{med}(\tilde{\omega})\}}{\tilde{\omega}-\omega_A}&
 \end{aligned}
\end{eqnarray}
which, rearranged, gives us the variant of the Kramers-Kronig relation (\ref{eq:Kramers-Kronig-special})
\begin{equation}
 \mathbb{P}\int_{-\infty}^\infty\!\!\!\! 
d\omega\frac{\omega^2}{c^2}\frac{\mathrm{Im}\{G_{ij}^{med}(\omega)\}}{\omega-\omega_A}=\pi\frac{\omega_A^2}{c^2}\mathrm{Re}\{G_{ij}^{med}(\omega_A)\}.
\end{equation}
Thus, one sees that the discussed term in (\ref{eq:dipoleshift_integralterm1}) is necessary in order to fulfill the Kramers-Kronig relations. 
Note that the integral term on the rhs of (\ref{eq:dipoleshift_integralterm2}) can be transformed into the one we got in (\ref{eq:dipoleshift_final}), making use of the well-known property of $\bar{\bar{G}}(\vec{r}_1,\vec{r}_2,\omega)$ to be real-valued for purely imaginary frequencies.

As we will see later on for the set-up discussed here and similar situations 
the second term in (\ref{eq:dipoleshift_integralterm1}) is indeed negligible
for emitter separations much larger than the characteristic length (i.e., the wavelength). However, when the emitters get close enough, the static ($\omega=0$) contribution in $\delta\omega_{12}$ increases substantially, and we cannot neglect the integral term any more. Also there may be situations such as the coupling
of quantum emitters over macroscopic distances through a negative-index material,
where this simple rule may not hold.

\section{A pair of emitters near a nano-wire}\label{subsec:a_pair_of_emitters_near_a_nano-wire}

\begin{figure}
\vspace*{0.5cm}
 \includegraphics[width=0.3\textwidth,angle=0]{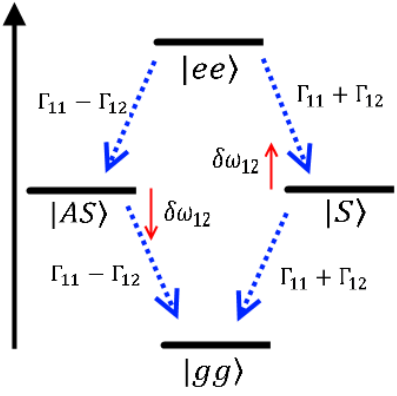}
\caption{\label{fig:dicke_levels} Collective atomic decays and dipole-dipole shifts of a pair of interacting two-level atoms, described in the Dicke basis. If there is a strong effective coupling (high $\Gamma_{12}$) between the atoms, the difference between the collective decay rates becomes large, leading to a superradiance phenomenon.}
\end{figure}

In the following, we will apply the method introduced above to a particular example. From the interaction of two emitters, each coupled to the basic surface-plasmon (SP) mode of a metallic wire of sub-wavelength radius, emerges a Dicke superradiance effect that depends on the inter-emitter distance (\cite{Dzsotjan-PRB-2010}). As seen in Fig.\ref{fig:dicke_levels}, the full collective atomic decays are $\Gamma_{11}\pm\Gamma_{12}$ and the singly excited levels get a $\pm\delta\omega_{12}$ wire-induced dipole-dipole shift.

Trying to calculate (\ref{eq:dipoleshift}) directly introduces difficulties since one has to deal with a principal value integral. This is especially a problem if we have to perform the integral numerically (which is usually the case by non-trivial geometries), because we have to know the behaviour of the $\bar{\bar{G}}^{med}(\omega)$ around $\omega=\omega_A$. As described in \cite{Dzsotjan-PRB-2010}, $\bar{\bar{G}}^{med}(\omega)=\int_0^{\infty}\!\! dk_z\tilde{\bar{\bar{G}}}^{med}(\omega;k_z)$, which is in this case  the scattered part of the Green's tensor. Although we know the analytic form of $\tilde{\bar{\bar{G}}}^{med}(k_z;\omega)$, it is a rather complicated function
and we cannot analytically integrate it. 
Using the method described in the previous section, however, circumvents these difficulties and lets us perform the much simpler integration in (\ref{eq:dipoleshift_final}) where, in addition we have to substitute purely imaginary frequencies into the Green's tensor, by which we get a more well-behaved function. 

\begin{figure}
 \includegraphics[width=0.5\textwidth,angle=0]{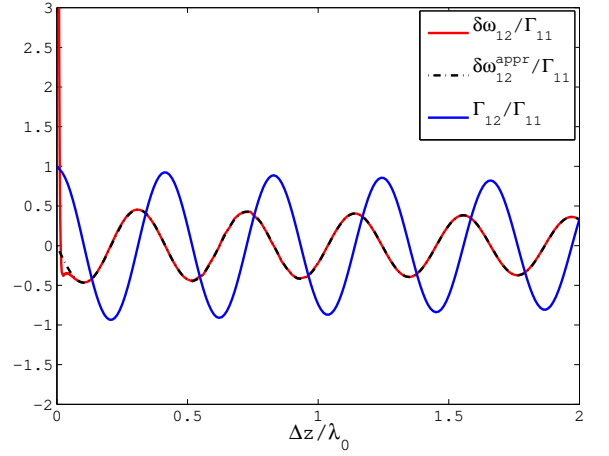}
\caption{\label{fig:dipoleshift} Dipole-dipole shift $\delta\omega_{12}$ due to the presence of the wire and the $\Gamma_{12}$ coupling between the atoms, scaled by the total single atom decay rate $\Gamma_{11}$, as a function of inter-atomic distance $\Delta z$ in units of the vacuum radiation wavelength $\lambda_0$. The wire radius is $0.01\lambda_0$ and the distance of the atoms from the wire axis is $0.015\lambda_0$. $\delta\omega_{12}^{appr}$ is a good analytic approximation of $\delta\omega_{12}$, for distances comparable to $\lambda_0$ and higher. Here, $\delta\omega_{12}/\Gamma_{11}$ shows an oscillating behaviour shifted by $\pi/2$ with respect to $\Gamma_{12}/\Gamma_{11}$. For inter-emitter distances comparable to the wire radius, there is a substantial increase in $\delta\omega_{12}$ and it begins to strongly deviate from $\delta\omega_{12}^{appr}$.}
\end{figure}

We calculated the dipole-dipole shift resulting from the presence of the wire using the full scattered Green's tensor. We compared it to an analytic approximation used by \cite{Gonzalez-Tudela-PRL-2011}, based on a single-plasmon resonance model,
a derivation of which is given in Appendix \ref{appendix:anal_approx}:
\begin{equation}
 \delta\omega_{12}^{appr}=-\frac{2\pi
 d^2\omega_A^2}{\hbar\epsilon_0c^2}A(\omega_A)\gamma(\omega_A)e^{-\gamma\Delta 
z}\sin(k_z^{pl}\Delta z).
\end{equation}
Here $A$ and $\gamma$ are the amplitude and width of the Lorentzian fit to the plasmonic resonances (see Appendix \ref{appendix:anal_approx})
and $k_z^{pl}$ is the longitudinal component of the wave vector of the plasmon mode. 
Fig. \ref{fig:dipoleshift} shows the results of the calculations. For inter-emitter distances larger than the vacuum radiaton wavelength, $\delta\omega_{12}^{appr}$ is a good approximation to the exact wire-induced dipole-dipole shift. It only deviates from the sinusoidal behaviour when the inter-emitter distance becomes comparable to the wire radius: in this case, the atoms begin to feel the three dimensional nature of the wire and the quasi-1D coupling approximation (i.e., $\delta\omega_{12}^{appr}$) is no longer valid. However, because the wire is quite thin, this typically happens at distances well below the vacuum radiation wavelength which means that in this regime the emitters are already strongly interacting through the vacuum as well. So we can safely say that the analytic approximation works well if the inter-emitter distance is above the vacuum radiation wavelength of the emitters.
In the regime where the exact calculations are well approximated by $\delta\omega_{12}^{appr}$, $\delta\omega_{12}/\Gamma_{11}$ oscillates with the same period as $\Gamma_{12}/\Gamma_{11}$ only with an additional $\pi/2$ relative phase shift \cite{Gonzalez-Tudela-PRL-2011}. This means that for inter-emitter distances yielding extrema for $\Gamma_{12}/\Gamma_{11}$, i.e., where the symmetric or antisymmetric transition is superradiant, $|S\rangle$ and $|AS\rangle$ are degenerate. On the other hand, when $\Gamma_{12}/\Gamma_{11}=0$, this degeneracy is lifted by $|\delta\omega_{12}|$ being maximal. The extrema of $|\delta\omega_{12}|$ are $0.5\Gamma_{11}$ at most. The decay of the amplitude of the oscillations for both  $\delta\omega_{12}/\Gamma_{11}$ and $\Gamma_{12}/\Gamma_{11}$ is caused by ohmic losses in the metal, represented by $\gamma$.
Thus, the interaction always makes a distinction between the symmetric and the antisymmetric transition: either by the different decay rates, or by the lifted degeneracy of $|S\rangle$ and $|AS\rangle$.

\begin{figure}
 \includegraphics[width=0.5\textwidth,angle=0]{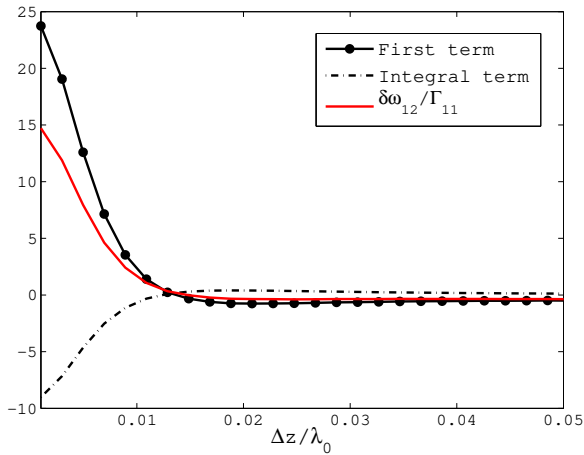}
\caption{\label{fig:partsofshift} The first term and the second, integral term of the dipole-dipole shift, as seen in (\ref{eq:dipoleshift_final}) and the actual wire-induced dipole-dipole shift $\delta\omega_{12}$, scaled by the total single-emitter decay rate $\Gamma_{11}$. The wire radius is $0.01\lambda_0$ and the distance of the atoms from the wire axis is $0.015\lambda_0$. Because of the increasing static ($\omega=0$) contributions in the Green's tensor, for small inter-emitter distances the integral term becomes comparable to the first term, becoming no longer negligible.}
\end{figure}

As seen in Fig.\ref{fig:partsofshift}, the closer the emitters are to each other, the more substantial the integral term of (\ref{eq:dipoleshift_final}) becomes. This is in accordance with the arguments made in Sec. \ref{sec:origin_of_int_term}, namely that the decrease of the inter-emitter distance enhances the static contribution of the Green's tensor. For small enough distances, the integral term is not negligible any more.

\section{Summary}\label{sec:summary}

In the present paper we discussed the effects of a tailored reservoir
on the Lamb and dipole-dipole shifts of quantum emitters coupled to a common
radiative reservoir. By considering only shifts relative to those in vacuum
all complications arizing from ultra-violet divergencies and off-resonant
contributions from other transitions were eliminated, reducing the problem
to the calculation of an integral of the electromagnetic Green tensor over
all frequencies. We presented a method that greatly simplifies this calculation 
by transforming the original expression containing a principal value integral over frequency into an ordinary integration over the imaginary axis.  The method does not imply any specific configuration or system, so it can be used in a wide variety of problems where level shifts due to dipole-dipole interaction have to be calculated. 
We discuss the appearance and importance of an integral term in the derived expression, that sweeps across purely imaginary frequencies, and is usually neglected in the literature.
We apply the method for calculating the dipole-dipole shift of a pair of atoms coupled to the guided surface plasmon modes of a metallic nano-wire and compare it to a quasi-1D analytic approximation. The results show that for inter-emitter distances comparable to the wire radius $\delta\omega_{12}$ becomes considerably larger than the single-atom decay rate ($\Gamma_{11}$) and the approximation doesn't hold. However, for larger distances the shift is very well approximated by the analytic expression, and $\delta\omega_{12}/\Gamma_{11}$ shows an oscillatory behaviour having roughly half the amplitude and the same period as $\Gamma_{12}/\Gamma_{11}$, as well as an additional phase shift of $\pi/2$. Thus, the interaction always makes a distinction between the symmetric and antisymmetric transition of the 2-atom system, either by the modified collective decay rates (superradiance) or the lifted degeneracy of $|S\rangle$ and $|AS\rangle$. We also take a look at the behaviour of the integral term mentioned above and conclude that it indeed becomes substantial for distances comparable to the wire radius.

\section*{Acknowledgements}

David Dzsotjan acknowledges financial support by the OPTIMAS Carl-Zeiss-PhD program and by the Research Fund of the Hungarian Academy of Sciences (OTKA) under contract No. NN 78112.

\section*{Appendix}

\begin{appendix}

\section{Kramers-Kronig relation generally}\label{appendix:Kramers-Kronig_relation_generally}
First, let us look at the Kramers-Kronig relations in case of a general, complex-valued function $f(\omega)$ which is analytic in the upper complex half-plane. According to Cauchy's theorem,
\begin{equation}
 f(\omega_A)=\mathrm{\lim_{\delta\rightarrow 0^+}}\frac{1}{2\pi\mathrm{i}}\varointctrclockwise_\mathrm{C}d\omega\frac{f(\omega)}{\omega-\omega_A-\mathrm{i}\delta}
\end{equation}
and the contour (containing the point $\omega_A+\mathrm{i}\delta$ within) is closed in the upper complex half-plane. If we assume that the path integral of $f(\omega)$ is nonzero only over the real axis and disappears on the other parts of the contour (that is, on the complex plane) if we extend it to infinity, we can write the integral as
\begin{equation}
 f(\omega_A)=\mathrm{\lim_{\delta\rightarrow 0^+}}\frac{1}{2\pi\mathrm{i}}\int_{-\infty}^{\infty}d\omega\frac{f(\omega)}{\omega-\omega_A-\mathrm{i}\delta}.
\end{equation}
Since
\begin{equation}
 \frac{1}{\omega-\omega_A-\mathrm{i}\delta}=\mathbb{P}\left(\frac{1}{\omega-\omega_A}\right)+\mathrm{i}\pi\delta(\omega-\omega_A),
\end{equation}
\begin{equation}
 f(\omega_A)=\frac{1}{2\pi\mathrm{i}}\mathbb{P}\int_{-\infty}^{\infty}d\omega\frac{f(\omega)}{\omega-\omega_A}+\frac{1}{2}f(\omega_A).
\end{equation}
And so:
\begin{equation}
 f(\omega_A)=\frac{1}{\pi\mathrm{i}}\mathbb{P}\int_{-\infty}^{\infty}d\omega\frac{f(\omega)}{\omega-\omega_A}
\end{equation}
or with real and imaginary parts
\begin{eqnarray}
\mathrm{Re}\{f(\omega_A)\}&=&\frac{1}{\pi}\mathbb{P}\int_{-\infty}^{\infty}\!\!\!\! d\omega\frac{\mathrm{Im}\{f(\omega)\}}{\omega-\omega_A}\nonumber\\
\mathrm{Im}\{f(\omega_A)\}&=&-\frac{1}{\pi}\mathbb{P}\int_{-\infty}^{\infty}\!\!\!\! d\omega\frac{\mathrm{Re}\{f(\omega)\}}{\omega-\omega_A}.
\end{eqnarray}

 \section{Analytic approximation of the dipole-dipole shift for two emitters coupled by a nano-wire}\label{appendix:anal_approx}

The Green's tensor for an infinitely long, cylindrical wire can be calculated as given in \cite{Dzsotjan-PRB-2010}. For the scattered part we can formally write
\begin{equation}
 \bar{\bar{G}}^{med}(\vec{r_1},\vec{r_2},\omega)=\int_{-\infty}^\infty\!\!\!\! dk_z\tilde{\bar{\bar{G}}}^{med}(\vec{r}_1,\vec{r}_2,\omega;k_z)
\end{equation}
where we know $\tilde{\bar{\bar{G}}}^{med}$ analytically. The atom-wire coupling is strongest if the dipole moment of the emitters point in the radial direction. If the cylindrical coordinates of two emitters only differ in their $z$ component (where $z$ is the symmetry axis of the wire) one finds
\begin{equation}\label{eq:green_cylinder_app}
 G_{rr}^{med}(\vec{r_1},\vec{r_2},\omega)=\int_{-\infty}^\infty\!\!\!\! dk_z e^{\mathrm{i}k_z\Delta z}\tilde{G}_{rr}^{med}(\vec{r}_1,\vec{r}_1,\omega;k_z)
\end{equation}
where $\Delta z=|z_2-z_1|$. In case of a thin wire (radius well below the vacuum radiation wavelength of the emitter), and small emitter-wire distance (in the order of magnitude of the radius), the plasmonic contribution becomes dominant in $\tilde{G}_{rr}^{med}(\vec{r_1},\vec{r_1},\omega;k_z)$. In this case, the imaginary part of the Green's tensor is well approximated by two Lorentzian fits, centered at $\pm k_z^{pl}$, i.e., the $z$ component of the wave vector of the plasmonic mode. This approximation is valid for inter-emitter distances larger than the vacuum radiation wavelength because in this case the only substantial channel that couples the emitters are the surface plasmon modes.
\begin{equation}
\begin{split}
\mathrm{Im}\{\tilde{G}_{rr}^{med} & (\vec{r}_1,\vec{r}_1,\omega;k_z)\}\approx \\
& \frac{A(\omega)}{1+\frac{[k_z-k_z^{pl}(\omega)]^2}{\gamma(\omega)^2}}+\frac{A(\omega)}{1+\frac{[k_z+k_z^{pl}(\omega)]^2}{\gamma(\omega)^2}}
\end{split}
\end{equation}
Because of the translational invariance of the wire in the $z$ direction, $\tilde{\bar{\bar{G}}}^{med}$ is symmetric in $k_z$. Thus, substituting in (\ref{eq:green_cylinder_app}) we get
\begin{equation}
 \mathrm{Im}\{G_{rr}^{med}(\vec{r}_1,\vec{r}_2,\omega)\}=\mathrm{Im}\{2\pi\mathrm{i}A\gamma e^{\mathrm{i}k_z^{pl}\Delta z}e^{-\gamma \Delta z}\}.
\end{equation}
\newline
With this, we can now express the wire-induced single-emitter decay rate and emitter-emitter coupling, respectively
\begin{eqnarray}
\Gamma_{11}^{appr}&=&\frac{4d^2\pi\omega_A^2}{\hbar\epsilon_0c^2}A(\omega_A)\gamma(\omega_A)\\
 \Gamma_{12}^{appr}&=&\frac{4d^2\pi\omega_A^2}{\hbar\epsilon_0c^2}A(\omega_A)\gamma(\omega_A)e^{-\gamma\Delta z}\cos(k_z^{pl}\Delta z),
\end{eqnarray}
as well as the wire-induced dipole-dipole shift, according to (\ref{eq:dipoleshift_final}):
\begin{equation}\label{eq:shift_wire_app}
\begin{aligned}
 \delta\omega_{12}^{appr}=&-\frac{2\pi d^2\omega_A^2}{\hbar\epsilon_0c^2}A(\omega_A)\gamma(\omega_A)e^{-\gamma\Delta z}\sin(k_z^{pl}\Delta z)\\
&+\frac{d^2\omega_A}{\hbar\epsilon_0\pi}\int_0^\infty\!\!\!\! d\kappa\frac{\kappa^2}{c^2}\frac{\mathrm{Re}\{2\pi\mathrm{i} A\gamma e^{-\gamma\Delta z}e^{\mathrm{i}k_z^{pl}\Delta z}\}_{\omega=\mathrm{i}\kappa}}{\kappa^2+\omega_A^2}.
\end{aligned}
\end{equation}
According to the discussion in the paper, for inter-emitter distances larger than the vacuum radiation wavelength the integral term in (\ref{eq:shift_wire_app}) can be neglected, so in the end we arrive to the analytic approximation
\begin{equation}
 \delta\omega_{12}^{appr}\approx-\frac{2\pi d^2\omega_A^2}{\hbar\epsilon_0c^2}A(\omega_A)\gamma(\omega_A)e^{-\gamma\Delta z}\sin(k_z^{pl}\Delta z).\\
\end{equation}
\end{appendix}

\end{document}